# A Proposal to Revise the Macroscopic Electromagnetic Theory

Jacob Neufeld

## 1. Introduction

The discussion which follows deals with a current widely used theory on wave propagation in dispersive and absorbing media.

The theory does not always lead to scientifically meaningful results. Inconsistencies and contradictions have been encountered. In fact, there are instances in which the theory becomes entirely useless. In those instances the widely used theory has broken down completely. It is unable to deal with simple observations in physics.

Documentary evidence is submitted in this discussion pointing out the seriousness of the situation.

A revised formulation is proposed in which the difficulties currently encountered do not occur. The proposed theory is based on the recognition that a distinction is made between an event which represents a cause and an event which represents the effect produced by the cause. As a consequence of the cause-effect relationship, a certain non-Maxwellian quantity which has been overlooked in the current widely used theory is introduced in the new formulation. The inclusion of such a non-Maxwellian quantity is essential in order to provide a meaningful physical system.

Effects of energy absorption have been analyzed in some of the referenced publications which appeared more than two decades ago [Neufeld 1969, 1970a,b]. In these accounts, absorption was treated as perturbation. In a different approach [Neufeld 1960], the effects of absorption are accounted for by certain non-Maxwellian quantities which have been referred to above.

It has been observed recently that the perturbation treatment needs to be reformulated. A study is now being made of the distinctive features of the perturbation theory as compared to the theory in which the non-Maxwellian quantities are introduced.

## 2. Revolutionary Findings by Ginzburg

Credit is due to Ginzburg [1964a] for certain revolutionary findings concerning the widely used theory on wave propagation in dispersive media. Ginzburg observed that "Despite the fact that the problem of the conservation law and the expression for the energy density in electrodynamics is a fundamental one, there are certain aspects of it which have not yet been elucidated, in particular for the case of an absorbing dispersive medium." Thus, according to Ginzburg, such a fundamental quantity as the mean electromagnetic energy density has been shown to be negative [Ginzburg 1964b]. Furthermore, he pointed out that "... when absorption is present, it is not in general possible to introduce phenomenologically the concept of the mean electromagnetic energy density" [Ginzburg 1964c].

The concept of energy density is not the only one discussed in the Ginzburg disclosure. He also considered the inadequacy of the current theory to account for the relationship between group velocity and the velocity of energy propagation in space. According to him, "When absorption is present, $v_{gr} = d\omega/dk$ is in general no longer meaningful and may, for example, give values exceeding the velocity of light in vacuum $c$..." [Ginzburg 1964d].

Ginzburg was fully aware of the seriousness of the situation. He admitted "... the author is aware that others besides himself have long been unclear concerning these matters" [Ginzburg 1964b]. Ginzburg did not suggest a solution to the perplexing problems.

It should be noted that the ideas of Ginzburg or related ideas of others deal with macroscopic as opposed to microscopic point of view.

# 3. Other Unrecognized Issues Apart from The Ginzburg Findings

The observations of Ginzburg are fully supported by certain findings of others. Some of these have been discussed in the above mentioned publications [Neufeld 1970a,b]. The following comments illustrate the current situation.

*Comment 1.* In the current theory the energy density of a system is not a uniquely defined quantity. There is considerable ambiguity on that account. According to the current theory, the energy of a system depends on the manner in which one arrives at a given state. In other words, energy density may have any value depending on the previous history of the system [Neufeld 1970c]. This is not acceptable, as flaws in logic should not be tolerated in a widely used theory.

*Comment 2.* There is a curious situation regarding the relationship between the expression for the energy density in an absorbing medium and the parameter which defines the process of absorption. It is generally understood in elementary mechanics that the energy of a system in which there is absorption does not depend on the parameter $\gamma$ which represents absorption.

For instance, the kinetic energy of a moving body is always $mv^2/2$ whether the body moves in a viscous medium or whether there is no viscosity. Viscosity is a parameter which represents energy dissipation and the velocity $v$ represents the state of the system. On the other hand, in the current theory this energy density depends on the parameter $\gamma$ [Neufeld 1970c].

*Comment 3.* Brillouin [1960] has shown that the equality of group velocity and the velocity of energy propagation is valid when the medium is non-absorbing. He was unable to extend his theory when there is absorption. He observed a "curious anomaly" in the absorption band. A difficulty appeared in an effort to establish a meaningful interpretation of *c/U* where *c* is the velocity of light and *U* is the velocity of energy propagation. According to his findings, "*c/U* can become less than 1, and even less than zero." Brillouin [1960] observed that in the region of absorption "... the group velocity no longer represents the velocity of a signal or of energy transport."  Brillouin did not provide an adequate explanation for the occurrence of the "curious anomaly."

It is also of interest to point out that the failure of the current theory reported by Brillouin has been independently pointed out by Ginzburg and reported above in this presentation.

*Comment 4.*  A closely related problem has been discussed In the *International Journal of Electronics* [Neufeld 1969].   It has been shown that when using the Maxwellian macroscopic approach and analyzing the dielectric constant, the concept of energy has not been properly incorporated in the kinetic plasma theory.  The difficulties are due to the Boltzmann collisional term which accounts for a change in the velocity distribution due to collisions alone.  If one attempts to replace the Boltzmann theory by the Maxwellian macroscopic formulation, one obtains an expression of energy density which is explicitly dependent on the collisional effect.  A difficulty arises similar to the one pointed out in comment 2.

## 4.  Unrecognized Urgency in Physical Sciences

There are reasons to believe that the current difficulties have not been sufficiently recognized by others.  Nevertheless, we are confronted with very critical and urgent issues.  These issues became urgent many years ago when Ginzburg introduced his revolutionary findings in electromagnetic theory.  They have remained urgent for several decades and are still urgent today.

## 5.  A Dilemma To Be Resolved

As illustrated above, there are instances in which the current theory is not even equipped to cope with the most fundamental problems in physics.  The scientific community  faces a dilemma on how to respond in a difficult situation.  Scientific prudence and plain common sense suggest that we have no choice. We have valid and well-supported reasons to reach a decision which, to our best knowledge, has never been suggested before. According to our best understanding, the standard theory which has been with us a number of years should be abandoned as it is inaccurate and misleading. A new approach should be explored. This is a drastic decision. However, it is the only scientifically based decision one can make.

The logic of this decision is unavoidable. Customarily, one example is sufficient to invalidate a theory if the example shows that the theory leads to physically unacceptable results. In our case, more than one example has been submitted by others.

As responsible scientists, we all wish to stress the importance of scientific prudence and of plain common sense as guiding principles in our undertakings. We find it extremely difficult to accept a theory with the understanding that the theory which we accept leads to meaningless results.

## 6. Supplementary Information On Electromagnetic Theory

The Maxwell equations per se do not always provide sufficient information when it comes to an analysis of dispersive media. In many instances, no information is provided regarding the structure of an atomic or molecular medium or plasma. To complete the information, supplementary relationships have been provided. A structural model is adopted in which the medium is represented by an assembly of electrons immersed in a continuously distributed positive charge and exposed to an incident wave. The response of a single electron in such an environment provides the necessary information. The supplementary relationships are combined with the Maxwell equation per se to provide a single self-construed and logical system for further exploration of the problem. A frame of reference is obtained which is then used to determine the dielectric constant and conductivity of the medium.

Consider the effects of energy absorption on wave propagation in dispersive media. If there is no absorption, the supplementary relationship can be expressed as follows:

$$m\ddot{r} + m\omega_b^2 r = F_1 , \qquad (1)$$

where

$$F_1 = eE . \qquad (2)$$

On the other hand, if the medium is absorbing, one has

$$m\ddot{r} + m\gamma\dot{r} + m\omega_b^2 r = F_2 . \qquad (3)$$

In the above relationships $m$ is the mass of an electron, $e$ is the charge of an electron, $E$ is the electric intensity, $r$ is the departure of an electron from its position of equilibrium, $\omega_b$ is the binding frequency, $\gamma$ is a parameter which represents absorption, $F_1$ is a driving force where there is no absorption, and $F_2$ is the driving force where there is absorption. An important quantity in the above expressions is represented by a frictional term $m\gamma\dot{r}$, which represents absorption.

## 7. An Ambiguity in the Interpretation of The Driving Force $F_2$

The supplementary relationships (1), (2), and (3) do not provide certain pertinent information regarding the driving force $F_2$. One needs to know whether or not the mechanism of energy absorption represented by $m\gamma\dot{r}$ has any effect on $F_2$. If it has no effect, then the driving force is the same whether there is absorption or not. In such case one has

$$F_1 = F_2. \tag{4}$$

On the other hand, if the mechanism of energy absorption has an effect on $F_2$, one has

$$F_1 \neq F_2. \tag{5}$$

There is a fundamental difference between the assumption that $F_1 = F_2$ and the assumption that $F_1 \neq F_2$.

The assumption (4) provides a framework for a theory which is widely accepted and widely used in our scientific undertakings.

The inequality $F_1 \neq F_2$ provides a framework for a formulation which is now proposed.

## 8. A Far Reaching Assumption Introduced By Lorentz

Reconsider the assumption that $F_1 = F_2$ and explore the consequences of this assumption. Having observed the inconsistencies reported by Ginzburg and others, the next step is to determine the roots of the difficulties. Consider an important happening which occurred about a century ago. At that time Lorentz [1916] introduced quite arbitrarily and without adequate justification certain assumptions which became a source of considerable difficulties in our scientific undertaking.  A situation was created which is of unique significance in the history of physics.

A considerable amount of scientific effort is now applied to ideas which are misguided.  Many scientific results which are considered as valid are based on a misleading theory.  It is surprising that critical appraisal of the Lorentz ideas are limited to the referenced publications.

Apparently, Lorentz overlooked an important aspect of the problem.  He did not observe that when there is absorption there must be a cause to be accounted for in making such an observation.  An entity which represents such a cause should be included in a meaningful formulation.

## 9.  A Cause and The Effect Produced By the Cause

The analysis leading to the new formulation is based on a strict observance of a principle dealing with a cause and the effect produced by the cause.  The cause-effect principle provides a logical link which connects an entity A which is the cause with an entity B which is the effect produced by the cause.

A simple example from elementary mechanics illustrates the situation.  When the speed of a moving body increases, the acceleration is the effect which is observed [entity B].  One then assumes that there exists a force which was applied to the body [entity A] which accounts for the observation.

There is a clear analogy between the effect associated with a moving body and the effect described in this presentation.  The application of the cause-effect principle gives a valuable insight which cannot be obtained otherwise.  By observing the cause-effect relationship, a logical link is provided which connects certain quantities to the exclusion of others.  In this analysis the application of the cause-effect principle has revealed that there are some essential quantities which have been overlooked in a complete description of the system.

## 10. A Non-Maxwellian Quantity Which Has Been Overlooked by Lorentz

Consider two fundamental processes or events which occur in the system. One of these deals with the storage of energy in space. The relevant term is $m\ddot{r} + m\omega^2 r$. It is used in the formulation of the dielectric constant of the medium interacting with the wave. The other process or event deals with energy dissipation. The relevant term is $m\gamma \dot{r}$. It is used in the determination of the conductivity.

Assume that there are two distinctive causes which account for the occurrence of these two fundamental, different events. One cause accounts for the storage of energy in space. The other cause deals with energy absorption. The cause of energy storage is expressed as $eE$. The relevant cause effect relationship is expressed as

$$eE = m\ddot{r} + m\omega_b^2 r. \tag{6}$$

In order to establish the cause-effect relationship for the process of energy absorption, a non-Maxwellian quantity is introduced which is designated as "impressed" force and represented symbolically as $eE^{(e)}$. The impressed force $eE^{(e)}$ is the cause of energy absorption. The relevant cause-effect relationship is expressed as

$$eE^{(e)} = m\gamma \dot{r}. \tag{7}$$

The relationship previously pointed out that $F_1 \neq F_2$ can now be established more precisely. It appears that the quantity $eF_2$ has two components. One of these deals with the energy storage in space. The other deals with energy absorption. One has

$$F_2 = eE + eE^{(e)}. \tag{8}$$

## 11. Comments by Abraham and by Becker

The ideas involving the cause-effect principle and the need for considering the non-Maxwellian impressed quantity have been clearly pointed out by Abraham [1930] and by Becker [1964]. Abraham observed that "...intensity **E** is not the only cause of the occurrence of current. There must also be other forces present, which tend to send a current through the conductor. We shall take account of these forces by introducing a vector $\mathbf{E}^{(e)}$, and extending Ohm's law to the form $i = \sigma(\mathbf{E} + \mathbf{E}^{(e)})$. For brevity we call $\mathbf{E}^{(e)}$ the *impressed* force', or *applied* force'... ." Furthermore, Abraham [1930] pointed out that "...the field of the impressed forces $\mathbf{E}^{(e)}$ being simply regarded as given, it will help to engender a more vivid apprehension of the subject if we consider briefly how that field arises in some special cases."

A similar statement supporting the significance of $\mathbf{E}^{(e)}$ has been made by Becker [1964]. According to Becker, "...we always started with the idea that the motion of charge carriers in conductors, and thus the flow of electric current, was produced solely by electrical field **E**. Now there are, however, other non-electrical causes by which a current can be made to flow through a conductor. We call such a cause an *impressed force*. If formally, this is expressed as the product $e\mathbf{E}^{(e)}$ of the carrier charge $e$ involved and an *impressed field strength* $\mathbf{E}^{(e)}$ herewith defined, we can take this field strength into account in Ohm's law, for example, through the expression $\mathbf{g} = \sigma(\mathbf{E} + \mathbf{E}^{(e)})$."

Both Abraham and Becker dealt with certain investigations in which non-electrical energy (chemical or mechanical) was transformed into electrical energy. In the current analysis the opposite effect is dealt with. Nevertheless, the problems are similar. The impressed forces used by Abraham and by Becker are analogous to the impressed forces in the current investigation.

## 12. Dielectric Constant and Conductivity According to Two Conflicting Theories

In summarizing, the supplementary relationships are arranged one beside the other as follows. According to Lorentz, if $\gamma = 0$ one has

$$m\ddot{r} + m\omega_b^2 r = eE . \tag{9}$$

On the other hand, if $\gamma \neq 0$, one has

$$m\ddot{r} + m\gamma\dot{r} + m\omega_b^2 r = eE . \tag{10}$$

The supplementary relationships are different in the proposed formulation. In the new formulation, one has

$$m\ddot{r} + m\omega_b^2 r = eE \tag{11}$$

and

$$m\gamma\dot{r} = eE^{(e)} . \tag{12}$$

The equality (11) is obtained by combining (3), (7), and (8).

In the proposed formulation there is a clear distinction between two fundamental processes in dispersive media. There is a process which leads to the formulation of the dielectric constant. It is expressed by (11). There is also the process which leads to the formulation of conductivity. It is expressed in (12). In the proposed formulation these two processes are independent. This means that any change in the mechanism which controls one of these processes does not affect the mechanism which controls the other process. On the other hand, according to Lorentz these processes are interdependent, as they appear in an expression such as (10).

Using supplementary relationships as a frame of reference, an expression for the dielectric constant and conductivity is obtained according to the two conflicting theories. Proceed at first with the theory of Lorentz.

Using the standard procedure and assuming that there is no absorption ($\gamma = 0$), one obtains an expression for the dielectric constant which is as follows:

$$\epsilon = 1 + \frac{\omega_0^2}{\omega_b^2 - \omega^2}, \tag{13}$$

where $\omega_0 = 4\pi Ne^2/n$ and $N$ is the number of electrons per unit volume. On the other hand, if there is absorption, one obtains

$$\epsilon = 1 + \frac{\omega_0^2}{\omega_b^2 - \omega^2 - i\gamma\omega} = \epsilon_1 + i\epsilon_2. \tag{14}$$

Thus, according to Lorentz [1916], the dielectric constant of an absorbing medium is a complex quantity which is dependent on the absorption parameter $\gamma$. In some instances it has been found convenient to assign to $\epsilon_1$ and to $\epsilon_2$ a particular physical meaning. It was assumed that $\epsilon_1$ represents an "effective" dielectric constant such that

$$\epsilon_{eff} = \epsilon_1 , \tag{15}$$

and that the conductivity $\sigma$ can be expressed as

$$\sigma = \frac{\omega \epsilon_2}{4\pi} . \tag{16}$$

A rationale supporting the validity of (16) has not been disclosed.

The next step is to determine the corresponding quantity according to the suggested formulation.

Using the standard procedure, one obtains from the supplementary relationship (11) an expression for the dielectric constant, as follows

$$\epsilon = 1 + \frac{\omega_0^2}{\omega_b^2 - \omega^2} . \tag{17}$$

In order to obtain an expression for conductivity in the new interpretation, one must determine the average rate of energy absorption in an oscillatory field having frequency $\omega$, as follows

$$Q = \overline{NeE^e \dot{r}} . \tag{18}$$

The bar above expression (18) represents an averaging process. The averaging extends over a time interval which is large when compared to $1/\omega$. The quantity $Q$ can also be expressed in terms of Ohm's law, i.e., one has

$$Q = \sigma E^2. \qquad (19)$$

Using the equalities (11), (12), (18), and (19), one obtains an expression for conductivity.

The results obtained according to two conflicting theories can be used as examples of very fundamental differences between the proposed formulation and that of Lorentz. According to the proposed formulation, the dielectric constant is always independent of the absorption parameter $\gamma$. In other words, the dielectric constant is always a real quantity whether there is absorption or not. Furthermore, a complex dielectric constant is not a valid scientific concept. It has no place in any physically acceptable scientific theory.

## 13. Concluding Remarks

Scientific undertakings during the last several decades were so outstanding and the achievements were so great, a situation developed in which certain problems in physics did not receive the attention which they deserved. It appears that an assumption was made that the electromagnetic theory as a branch of physics and technology was complete and no fundamental principle deserved further exploration. Such an assumption is not acceptable. There are serious flaws in the traditional well-accepted theory and therefore fundamental changes are required. The proposed formulation offers a valid alternative solution.

**References**

Abraham Max, *The Classical Theory of Electricity and Magnetism,* [revised by Richard Becker (New York, NY: Hafner Publishing Co., II Edition, 1930) pp. 116-117].

Becker Richard, *Electromagnetic Fields and Interactions*; Vol. I, Electromagnetic Theory and Relativity, [edited by Fritz Sauter (New York, NY: Blaisdel Publishing Co., 1964) p. 149].

Brillouin Leon, *Wave Propagation and Group Velocity* (New York, NY: Academic Press, 1960) p. 122.

Ginzburg, V L, *The Propagation of Electromagnetic Waves in Plasmas* (Oxford: Pergamon Press, 1964a).

Ginzburg V L, *The Propagation of Electromagnetic Waves in Plasmas* (Oxford: Pergamon Press, 1964b) p. 477.

Ginzburg V L, *The Propagation of Electromagnetic Waves in Plasmas* (Oxford: Pergamon Press, 1964c) p. 244.

Ginzburg V L, *The Propagation of Electromagnetic Waves in Plasmas* (Oxford: Pergamon Press, 1964d) p. 249.

Lorentz H A, *The Theory of Electrons and Its Application to the Phenomena of Light and Radiant Heat,* A course of lectures delivered at Columbia University, New York, NY, March-April 1906 (Leipzig, Germany: II Edition, 1916) p. 132.

Neufeld Jacob, *Phys. Rev.* **152**, No. 2, 708-717 (1960).

Neufeld Jacob, *Int. J. Electronics* **27** 301-317 (1969).

Neufeld Jacob, *Il Nuovo Cimento* **X65B** 33-69 (1970a).

Neufeld Jacob, *Il Nuovo Cimento* **X66B** 51-76 (1970b).

Neufeld Jacob, *Il Nuovo Cimento* **X65B** 55 (1970c).